\documentstyle[11pt]{article}
\input{psfig.sty}
\def \SAIT #1 #2 {{\em Mem.\ Soc.\ Astron.\ It.\/} {\bf #1}, #2}
\def \MESS #1 #2 {{\em The Messenger\/} {\bf #1}, #2}
\def \ASTRNACH #1 #2 {{\em Astron. Nach.\/} {\bf #1}, #2}
\def \AAP #1 #2 {{\em Astron. Astrophys.\/} {\bf #1}, #2}
\def \AAL #1 #2 {{\em Astron. Astrophys. Lett.\/} {\bf #1}, L#2}
\def \AAR #1 #2 {{\em Astron. Astrophys. Rev.\/} {\bf #1}, #2}
\def \AAS #1 #2 {{\em Astron. Astrophys. Suppl. Ser.\/} {\bf #1}, #2}
\def \AJ #1 #2 {{\em Astron. J.\/} {\bf #1}, #2}
\def \ANNREV #1 #2 {{\em Ann. Rev. Astron. Astrophys.\/} {\bf #1}, #2}
\def \APJ #1 #2 {{\em Astrophys. J.\/} {\bf #1}, #2}
\def \APJL #1 #2 {{\em Astrophys. J. Lett.\/} {\bf #1}, L#2}
\def \APJS #1 #2 {{\em Astrophys. J. Suppl.\/} {\bf #1}, #2}
\def \APSS #1 #2 {{\em Astrophys. Space Sci.\/} {\bf #1}, #2}
\def \ASR #1 #2 {{\em Adv. Space Res.\/} {\bf #1}, #2}
\def \BAIC #1 #2 {{\em Bull. Astron. Inst. Czechosl.\/} {\bf #1}, #2}
\def \JSQRT #1 #2 {{\em J. Quant. Spectrosc. Radiat. Transfer\/} {\bf #1}, #2}
\def \MN #1 #2 {{\em Mon. Not. R. Astr. Soc.\/} {\bf #1}, #2}
\def \MEM #1 #2 {{\em Mem. R. Astr. Soc.\/} {\bf #1}, #2}
\def \PLR #1 #2 {{\em Phys. Lett. Rev.\/} {\bf #1}, #2}
\def \PASJ #1 #2 {{\em Publ. Astron. Soc. Japan\/} {\bf #1}, #2}
\def \PASP #1 #2 {{\em Publ. Astr. Soc. Pacific\/} {\bf #1}, #2}
\def \NAT #1 #2 {{\em Nature\/} {\bf #1}, #2}
\begin{document}
\normalsize

\large

\centerline{\bf {SPECIAL RELATIVITY AT ACTION IN THE UNIVERSE}}

\normalsize

\begin{center}
{Gabriele Ghisellini }

{\it Osservatorio Astronomico di Brera, v. Bianchi 46, I-23807 Merate, 
Italy\\  e-mail address: gabriele@merate.mi.astro.it}
\end{center} 

\centerline{\bf Abstract}
\noindent
Nature succeeds in accelerating extended and massive objects to 
relativistic velocities.
Jets in Active Galactic Nuclei and in galactic superluminal
sources and gamma--ray bursts fireballs have bulk Lorentz factors
from a few to several hundreds.
A variety of effects then arises, such as the beaming of the produced
radiation, light aberration, time contraction and the Doppler frequency shift.
I will emphasize that special relativity applied to real (i.e. extended) observed 
objects inevitably must take into account that any piece of information is carried by
photons. 
Being created in different parts of the source, they travel different
paths to reach the observer, depending on the viewing angle.
The object is seen rotated, not contracted, and at small viewing angles
time intervals are observed shorter than intrinsic ones.

\normalsize

\section{Introduction}
We are used to consider earth's particle accelerators as the realm 
of special relativity.
But nature can do better and bigger, and is able to accelerate
really large masses to relativistic speeds.
Besides cosmic rays, we now know that in the jets of active galactic nuclei
(AGN) which are also strong radio emitters plasma flows at $v\sim 0.99c$,
and that in gamma--ray bursts (GRB) the fireball resulting from the release
of $\sim 10^{52}$ erg in a volume of radius comparable to the
Schwarzschild radius of a solar mass black hole reaches $v\sim 0.999c$.
These speeds are so close to the light speed that it is more convenient
to use the corresponding Lorentz factor of the bulk motion,
$\Gamma = (1-\beta^2)^{-1/2}$: for $\Gamma \gg 1$, we have
$\beta \sim 1-1/(2\Gamma^2)$.

The energetics involved is huge.
In AGNs, a fraction of a solar mass per year can be accelerated to 
$\Gamma\sim 10$, leading to powers of $\sim 10^{46}$ erg s$^{-1}$ in
bulk motion.
In GRBs, the radiation we see, if isotropically emitted,
can reach $10^{54}$ erg s$^{-1}$, suggesting even larger values
for the bulk motion power.
Recently, very interesting sources have been discovered within our Galaxy 
through their activity in X--rays: they occasionally produce radio jets
closely resembling those of radio--loud quasars.
During these ejection episodes, the power in bulk motion can exceed
$10^{40}$ erg s$^{-1}$, a value thought to exceed the Eddington
limit for these sources.

The study of these $extended$ objects moving close to $c$ requires
to take into account the different travel paths of the photons reaching us.
Curiously enough, the resulting effects had not been studied until 1959, when
% Penrose (1959) and 
Terrel (1959) 
% independently 
pointed out that a
moving sphere does not $appear$ contracted, but $rotated$, contrary
to what was generally thought (even by Einstein himself ...).
These results, which were ``academic" in those years, are now fully
applied to the above mentioned relativistic cosmic objects.

In this paper I will present some of the evidence in support of
relativistic bulk motion in astronomical objects, and then discuss 
how ``text--book special relativity" has to be applied when information 
are carried by photons.

\section{Superluminal motion}
Rees (1966) realized that an efficient way of transporting energy
from the vicinity of super-massive black holes to the radio lobes of
the recently discovered radiogalaxies is 
through the bulk motion of relativistically moving plasma.
If this plasma emits radiation on its way, then we ought to see 
moving spots of emission in radio maps.
One of the most spectacular prediction by Rees was that  
this motion could appear to exceed the speed of light.
This was indeed confirmed in the early seventies, 
when the radio--interferometric techniques
allowed to link radio telescopes thousands of kilometers apart.
Among the first few observed targets was 3C 279, a radio--loud quasar
at a redshift of $z=0.538$.
Bright spots in radio maps taken at interval of months were apparently
moving at a speed exceeding 10 times $c$.
For obvious reasons, sources presenting this phenomenon are referred to as
{\it superluminal sources}.

This phenomenon can be simply explained, as long as the plasma is moving at 
velocities close to $c$ at small viewing angle (i.e. the angle
between the velocity vector and the observer's line of sight).
Consider Fig. 1: suppose the moving blob emits a photon
from positions $A$ and then from $B$. 
\begin{figure}
\vskip -1 true cm
\psfig{file=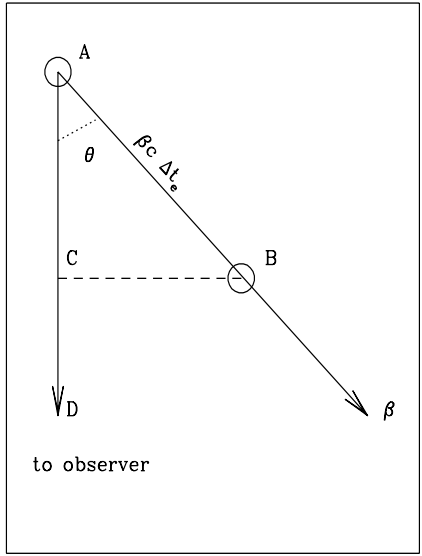,width=9truecm,height=10truecm}
\vskip -2 true cm
\caption[h]{Explanation of the apparent superluminal speed. 
A blob emits a photon from $A$ and, after a time $\Delta t_e$, from $B$.
If the true velocity is close to the speed of light and  the angle
$\theta$ is sufficiently small, the apparent velocity will exceed
the speed of light.}
\end{figure}
The time between the two emissions, as measured by an observer which sees 
the blob moving, is $\Delta t_e$.
Therefore the distance $AB$ is equal to $\beta c \Delta t_e$,
and $AC=\beta c \Delta t_e\cos\theta$.
In the same time interval, the photon emitted in $A$ has reached the point
$D$, and the distance $AD$ is equal to $c\Delta t_e$.
Thus the two photons are now separated by $DC=AD-AC=
c\Delta t _e(1-\beta\cos\theta)$.
The difference between the arrival times of these two photons
is then $\Delta t_a=\Delta t_e (1-\beta\cos\theta)$, and
the projected separation of the blob in the two images is
$CB=c \beta \Delta t_e \sin\theta$, leading to an apparent
velocity:
\begin{equation}
\beta_{app} \, =\, {\beta \sin\theta \over 1-\beta\cos\theta}
\end{equation}
It can be readily seen that $\beta_{app}>1$ for $\beta\to 1$ and 
small viewing angles $\theta$.
The apparent speed is maximized for $\cos\theta=\beta$, where
$\beta_{app}=\beta\Gamma$.
Notice that this simple derivation {\it does not require} any
Lorentz transformation (no $\Gamma$ factor involved!).
The superluminal effect arises 
only from the Doppler contraction of the arrival
times of photons.

\section{Beaming}

\noindent
Let us assume that a source emits isotropically in its rest frame $K^\prime$.
In the frame $K$, where the source is moving relativistically, 
the radiation is strongly anisotropic, and three effects occur:
\begin{itemize}
\item {\bf Light aberration:} photons emitted
at right angles
with respect to the velocity vector (in $K^\prime$) are 
observed in $K$ to make an angle given by
$\sin\theta =1/\Gamma$.
This means that in $K$ half of the photons are concentrated in a cone
of semi-aperture angle corresponding to $\sin\theta =1/\Gamma$.
\item {\bf Arrival time of the photons}: as discussed above, the emission and 
arrival time intervals are different.
As measured in the same frame $K$ we have, as before, 
$\Delta t_a =\Delta t_e (1-\beta\cos\theta)$.
If $\Delta t_e^\prime$ is measured in $K^\prime$,  
$\Delta t_e =\Gamma \Delta t_e^\prime$ leading to
\begin{equation}
\Delta t_a\, = \, \Gamma (1-\beta\cos\theta ) \Delta t_e^\prime \, 
\equiv \, {\Delta t_e^\prime \over \delta}
\end{equation}
Here we have introduced the factor $\delta$, referred to as the beaming
or Doppler factor. It exceeds unity for small viewing angles, and if so, 
observed time intervals are {\it contracted}.
\item {\bf Blueshift/Redshift of frequencies}: since frequencies are the
inverse of times, we just have $\nu =\delta \nu^\prime$.
\end{itemize}
\noindent
It can be demonstrated (see e.g. Rybicki \& Lightman 1979)
that the specific intensity $I(\nu)$ divided by the cube of the frequency
is Lorentz invariant, and therefore
\begin{equation}
I(\nu) \, =\, \delta^3 I^\prime(\nu^\prime)\, =\,
\delta^3 I^\prime(\nu/\delta).
\end{equation}
Integration over frequencies yields $I=\delta^4 I^\prime$.
The corresponding transformation between bolometric {\it luminosities} 
sometimes generates confusion.
It is often said that $L=\delta^4 L^\prime$.
What this means is: if we estimate the observed luminosity $L$ 
from the received flux {\it under the assumption of isotropy},  
this is related to $L^\prime$ through the above equation.

But suppose that the photon receiver covers the entire sky, i.e. completely
surrounds the emitting source: what is then the relation between 
the received power in the frames $K$ and $K^\prime$?
In this case we must integrate $dL/d\Omega$ over the solid angle obtaining:
\begin{equation}
L\, =\, \int \delta^4 {L^\prime \over 4 \pi} 2\pi \sin\theta  d\theta \,=\,
\Gamma^2 {\beta^2 + 3 \over 3} L^\prime \, \to \, {4\over 3} \Gamma^2 L^\prime
\end{equation}
Note that in this situation {\it the power is not a Lorentz invariant}.
This is because here we are concerned with the power {\it received} in the
two frames, not with the {\it emitted} one (which is indeed
Lorentz invariant). 
This is yet another difference with respect to 
``text--book" special relativity not accounting for  
photons: here the time transformation involves the Doppler
term $(1-\beta\cos\theta)$, causing the difference between
{\it emitted} and {\it received} power (see also Rybicki \& Lightman
1979, p. 141).

Because of beaming, relativistically moving objects appear much brighter if 
their beams point at us, and can therefore be visible up to large distances.
Besides being extremely important in order to calculate the intrinsic physical
parameters of a moving source, beaming is also crucial for the moving
object itself. The observed objects are rarely isolated and more often 
are part of a jet immersed in a bath of radiation. 
Just for illustration, let us consider a blob moving
close to an accretion disk and surrounded by gas clouds responsible
for the emission of the broad lines seen in the spectra of quasars: 
as the blob moves at relativistic speed in a bath of photons it 
will see this radiation enhanced. Furthermore, 
because of aberration, in its frame 
most of the photons will appear to come from the hemisphere
towards which the blob is moving, and be blueshifted.
For an observer at rest with respect to the photon bath,
the radiation energy density seen by the blob is enhanced by $\sim \Gamma^2$.
This increases the rate of interaction between the photons and the
electrons in the blob, leading to enhanced inverse Compton emission and possibly 
even deceleration of the blob by the so called {\it Compton drag effect}.

\section{Evidences for relativistic motion}
\subsection{Radio--loud AGN with flat radio spectrum}

\noindent
$\bullet$ {\bf Superluminal motion ---}
The most striking evidence of bulk motion comes from 
the observation of superluminal sources.
Improvements in the interferometric techniques have led to the discovery
of more than 100 of these sources (see e.g. Vermeulen \& Cohen 1994).
The typical bulk Lorentz factors inferred range between 5 and 20.

\vskip 0.25 true cm
\noindent
$\bullet$ {\bf Compton emission ---} From radio data (size, flux, spectrum) 
one can derive, through the
synchrotron theory, the number density of emitting particles
and the radiation energy density (Hoyle, Burbidge \& Sargent 1966).
These quantities determine the probability that particles and photons
interact through the inverse Compton process, and thus 
it is possible to predict the amount of the high energy radiation 
(i.e. X--rays) produced.
However, this estimated flux 
is often orders of magnitude larger than what is observed, if beaming
is not taken into account.
Conversely, the 
requirement that the radio source emits at most the observed X--ray flux,
sets a (lower) limit on the beaming factor $\delta$.
Typical values are in agreement with those derived from superluminal motion 
(Ghisellini et al. 1993).

\vskip 0.25 true cm
\noindent
$\bullet$ {\bf High brightness temperatures ---} 
This argument is similar to that just presented. 
The brightness temperature $T_b$ 
[defined through $I(\nu)\equiv 2kT_b\nu^2/c^2$]
is related to the density of particles and photons, and therefore
to the probability of Compton scattering: 
high brightness temperature implies powerful Compton emission.
More precisely, if $T_b>10^{12}$ K (this value is called the {\it Compton limit}), 
the luminosity produced by the first order Compton 
scattering is larger than the synchrotron luminosity, and 
that in the second order exceeds (by the same factor)
that in the first order, and so on. Clearly, this can only occur until 
the typical photon energy 
reaches the typical electron energy,  
above which the power has to drop.
This increasingly important particle cooling is called the {\it Compton catastrophe}.
To avoid it, we resort to beaming, recalling that $T_b$ transforms
according to
\begin{eqnarray}
T_b \, &=& \, \delta T_b^\prime\quad {\rm Source~ size~measured~ directly}
\nonumber \\
T_b\, &=& \, \delta^3 T_b^\prime\quad {\rm Source~ size~ measured~ through~ 
variability}
\end{eqnarray}
The Compton limit of $T_b>10^{12}$ K is derived by requiring that the radiation energy
density is smaller than the magnetic energy density.
A more severe limit 
can be obtained imposing the condition of equipartition between 
particle and magnetic energy densities, as proposed by Readhead (1994).
In the latter case one derives larger $\delta$ factors,
which are in agreement with those obtained by the two previous methods.
A note of caution: there is a significant number of sources, called {\it intraday 
variables}, whose radio flux changes on a timescale of {\it hours}.
For them, $T_b>10^{18}$ K, and therefore it 
 is inferred
 $\delta > 100$, a value too large
to be consistent with the those derived in other ways.
This is an open issue, and probably effects due to 
interstellar scintillation and/or coherent radiation (as in pulsars)
have to been invoked
(for a review see  Wagner \& Witzel 1995).

\vskip 0.25 true cm
\noindent
$\bullet$ {\bf Gamma--ray emission ---}
The $\gamma$--ray satellite CGRO discovered that radio--loud quasars
with flat radio spectrum (FSRQ) can be strong $\gamma$--ray emitters, 
and that often  most of their power is radiated in this band.
This emission is also strongly variable, on timescales of days or less.

However, photons above $m_ec^2=$511 keV can interact with lower energy
photons, producing electron--positron pairs.
This happens if the optical depth for the photon--photon interaction
is greater than unity, and this depends on the density
of the target photons and the typical size of the region
they occupy.
If not beamed, the large power and rapid variability observed imply
optical depths largely in excess of unity, and so $\gamma$--ray would be 
absorbed within the source. 
The condition of transparency to this process leads to lower limits on 
beaming factors somewhat smaller than in the previous cases.
There are also objects (still a few, but increasing in number) 
observed above a few tenths of a TeV (from the ground by Cherenkov 
telescopes). The corresponding limits on $\delta$ are the most severe.

\vskip 0.25 true cm
\noindent
$\bullet$ {\bf One--sidedness of jets ---}
Radio--sources are very often characterized by two lobes of extended
emission, but in some of them only one jet - starting from the nucleus
and pointing to one of the lobes - is visible.
If the emitting plasma 
is relativistically moving, the radiation from the 
jet approaching us is enhanced, while that from the receding jet
is dimmed, explaining the observed asymmetry.
This is often referred to as {\it Doppler favoritism}.

\vskip 0.25 true cm
\noindent
$\bullet$ {\bf Super--Eddington luminosities ---}
FSRQs show violent activity: large luminosity changes on short timescales.
If the variability timescales are associated 
to the Schwarzschild radius (hence the
black hole mass) luminosities 
exceeding the Eddington limit are often derived (if isotropy is assumed).
This difficulty can be easily overcome by beaming (which affects both the variability
timescales and the observed power).

\vskip 0.25 true cm
\noindent 
$\bullet$ {\bf The Laing--Garrington effect ---}
The two lobes of radio sources are often differently polarized
(or, more precisely, they are differently depolarized):
as always the (one sided) jet points towards the more
polarized lobe, this convincingly  
indicates that the visible jet is the one approaching the observer
(Laing 1988, Garrington et al. 1988).

\vskip 0.25 true cm
\noindent
$\bullet$ {\bf Jet bending ---} 
Jets of AGNs are often significantly  curved, posing problems of stability.
But the strong bendings could be largely apparent, if the jet is
seen at small viewing angles.
This constitutes a further independent hint
that these jets are pointing at us (although is not direct 
evidence of relativistic motion).

\vskip 0.25 true cm
\noindent
$\bullet$ {\bf Parent population ---} 
For each source whose jet is direct towards us, 
there must be $\sim\Gamma^2$ other sources with jets pointing away.
If beaming is important, then these objects 
must be less luminous, not (extremely) superluminal, and not showing violent
activity. This {\it parent population} of sources can be identified with 
that of 
radiogalaxies (Blandford \& Rees 1978; for a review see Urry \& Padovani 1995): 
indeed their number agrees with what expected from the beaming 
parameters derived by the other methods. 

\subsection{Galactic superluminal sources}

\noindent
In 1994, the two X--ray transients GRS1915+105
and J1655--40 were monitored with the VLA during a campaign 
aimed at finding radio jets associated with galactic sources.
Surprisingly, it was discovered superluminal motion
in both of them (Mirabel \& Rodriguez 1994; Hjellming \& Rupen 1995).
The most interesting observational fact is that in these 
two sources we see {\it both} the jet and the counter-jet.
This is unprecedented among superluminal sources, and it is possible because:
i) the viewing angle is large, suppressing the Doppler favoritism effect;
ii) the bulk Lorentz factor is small, $\Gamma=2.5$, leading to
a moderate beaming and thus an observable flux even at large viewing angles.

Under the assumption that the superluminal blobs  
move in 
opposite directions and are characterized by the same (true) velocity $\beta c$,
we can apply Eq. (1) $twice$ to derive $both$ $\beta$ and $\theta$.
This in turn determines $\delta$.

Very severe limits can be set on the power 
carried by the moving blob in the form of both particle bulk motion
and Poynting flux.
The radio emission observed with VLA comes from a region 
$\sim 10^{15}$ cm in size, and is believed to be produced by synchrotron
emission:  it is then possible to calculate the number of particles 
responsible for the observed radiation (this is a lower limit, since other 
non emitting,  i.e. sub relativistic, 
particles might be present).

The number 
inferred 
depends on the value of the magnetic field $B$:
an increase of $B$ decreases the amount of particles required, but
increases the implied Poynting flux.
{\it The sum of the power in particle bulk motion 
and in Poynting flux has therefore  a minimum.}
For GRS1915+105, this is of the order of $10^{40}$ erg s$^{-1}$ 
(Gliozzi, Bodo \& Ghisellini, 1999), and exceeds by a factor $\sim$ 10 the power emitted
by the accretion disk (in soft and medium energy X--rays).
This result itself puts strong constraints on any model for the 
acceleration of jets, by excluding the possibility that this occurs through
radiation pressure. 

\subsection{Gamma--ray Bursts}

\noindent
GRBs are flashes of hard X-- and $\gamma$--rays, lasting for a few seconds.
Discovered by the Vela satellites in the late sixties, their origin
remained mysterious until the Italian--Dutch satellite {\it Beppo}SAX
succeeded in locating them on the sky with small enough error--boxes:
prompt follow--up observations were then possible and led to the discovery 
that they are at cosmological distances.
This in turn allowed to estimate the energy involved which, 
if the radiation is emitted isotropically, 
is in the range $10^{52}$--$10^{54}$ erg. 
The energetics is in itself a strong evidence of relativistic bulk motion:
the short duration and the even shorter variability timescale
(of the order of 1 ms) imply a huge compactness (i.e. the luminosity
over size ratio). 
This resembles the conditions during the Big Bang, and implies a similar 
evolution: no matter in which form the energy is initially injected,
a quasi--thermal equilibrium between matter and radiation is reached,
with the formation of electron--positron pairs accelerated to
relativistic speeds by the high internal pressure.
This is a {\it fireball}.
When the temperature of the radiation (as measured in the comoving
frame) drops below $\sim$50 keV the pairs annihilate 
faster then the rate at which are produced
(50 keV, not 511 keV, as a thermal photon distribution has a
high energy tail...).
But the presence of even a small amount of barions, corresponding to 
only $\sim 10^{-6}~ M_\odot$,
makes the fireball opaque to Thomson scattering: the internal radiation
thus continues to accelerate the fireball until most of its initial 
energy has been converted into bulk motion. 
The fireball then expands at a constant speed and at some point 
becomes transparent.
If the central engine is not completely impulsive, but works intermittently,
it can produce many shells (i.e. many fireballs)
with slightly different Lorentz factors.
Late but faster shells can catch up early slower ones,
producing shocks which give rise to the observed burst emission.
In the meantime, all shells interact with the interstellar medium, and 
at some point the amount of swept up matter is large enough to decelerate
the fireball and produce other radiation which can be identified with 
the afterglow emission observed at all frequencies.

For GRBs the limits to the required bulk Lorentz factors follow
mainly from two arguments (for reviews see Piran 1999; Meszaros 1999):

\vskip 0.25 true cm
\noindent
$\bullet$ {\bf Compactness ---} We see high energy (i.e. $\gamma$--rays
above 100 MeV) emission, varying on short timescales.
As for the AGN case (see above) bulk motion is required 
to lower the implied luminosity and increase the intrinsic 
size of the emission region in order to decrease the opacity to pair 
production.
Limits to the Lorentz factor in the range 100--300 are derived.

\vskip 0.25 true cm
\noindent
$\bullet$ {\bf Variability ---} 
The radiation we see originates when 
the fireball has become 
optically thin to Thomson scattering, 
i.e. when it has expanded to radii of the order of $R_t=10^{13}$ cm.
Yet, we see 1 ms variability timescales.
Assuming that the radiation is produced while the fireball expands
a factor two in radius, short timescales are possible if
$\Delta t_a = (1-\beta\cos\theta)R_t/c$. Furthermore, 
if isotropic, we always see the portion of it 
pointing at us ($\theta=0^\circ$), 
and thus $\Delta t_a\sim 0.01 (R_t/10^{13})(100/\Gamma)^2$
Also this argument leads to values of $\Gamma$ in the range 100--300.

\section{Rulers and clocks vs photographs and light curves}

\noindent
In special relativity we are used to two fundamentals effects:
\begin{itemize}
\item Lengths shrink in the direction of motion
\item Times get longer
\end{itemize}
\noindent
The Lorentz transformations for a motion along the 
$x$ axis are ($K$ is the lab frame and $K^\prime$ is the comoving one) 
\begin{eqnarray}
x^\prime \, &=& \, \Gamma(x-vt)\nonumber \\ 
t^\prime\, &=& \,  \Gamma\left( 1- {v\over c^2} x\right) 
\end{eqnarray}
with the inverse relations given by
\begin{eqnarray}
x\, &=& \, \Gamma(x^\prime + vt^\prime)  \nonumber \\
t\, &=& \, \Gamma\left( t^\prime + {v\over c^2} x^\prime\right). 
\end{eqnarray}
The length of a moving ruler has to be measured 
through the position of its extremes {\it at the same time $t$}.
Therefore, as $\Delta t=0$, we have 
\begin{equation}
x^\prime_2 -x^\prime_1 \, =\, 
\Gamma(x_2-x_1)-\Gamma v\Delta t \, =\,
\Gamma(x_2-x_1)
\end{equation}
i.e. 
\begin{equation}
\Delta x \,=\,  {\Delta x^\prime \over \Gamma }
\, \to \, {\rm contraction}
\end{equation}
Similarly in order to determine a time interval 
a  (lab) clock has to be compared 
with one in the comoving frame,
which has, in this frame, {\it the same position $x^\prime$}.
It follows 
\begin{equation}
\Delta t \, =\, \Gamma \Delta t^\prime + \Gamma {v \over c^2} 
\Delta x^\prime
\, =\, 
\Gamma \Delta t^\prime \, \to \, {\rm dilation}
\end{equation}
An easy way to remember the transformations is to think to mesons
produced in collisions of cosmic rays in the high atmosphere, which 
can be detected even is their lifetime
(in the comoving frame) is much shorter than the time
needed to reach the earth's surface.
For us, on ground, relativistic mesons live longer.

All this is correct if we measure lengths by comparing
rulers (at the same time) and by comparing clocks (at the same position)
- the meson lifetime $is$ a clock.
In other words, {\bf if we do not use photons}
for the measurement process.

\subsection{The moving bar}

\begin{figure}
\vskip -1 true cm
\psfig{file=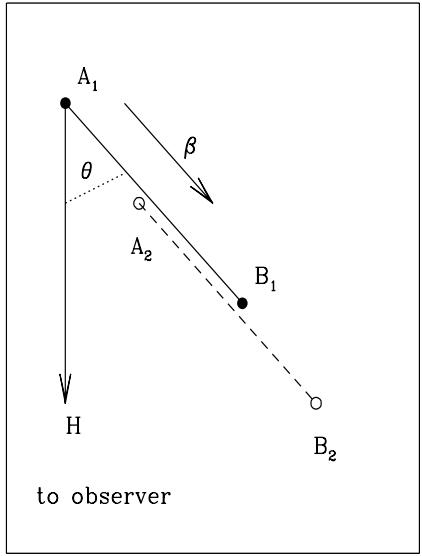,width=9truecm,height=10truecm}
\vskip -1 true cm
\caption[h]{A bar moving with velocity $\beta c$ in the direction of
its length. The path of the photons emitted by the extreme $A$ is longer
than the path of photons emitted by $B$.
When we make a picture of the bar (or a map), we collect photons
reaching the detector simultaneously . 
Therefore the photons from $A$ have to be emitted before those
from $B$,
when the bar occupied another position.}
\end{figure}

If the information (about position and time)
are carried by photons, we {\bf must} take into account
their (different) paths.
When we take a picture, we detect photons arriving at the same
time to our camera: if the moving body which emitted them is 
extended, we must consider that these photons have been emitted
at different times, when the moving object occupied different
locations in space.
This may seem quite obvious. And it is. Nevertheless
these facts were pointed out in 1959 (Terrel 1959), more than 50 years
after the publication of the theory of special relativity.

Let us consider a moving bar, of proper dimension $\ell^\prime$, moving 
in the direction of its length at velocity $\beta c$ and 
at an angle $\theta$ with respect to the line of sight (see Fig. 2).
The length of the bar in the frame $K$ (according to relativity
``without photons") is $\ell =\ell^\prime/\Gamma$.
The photon emitted in $A_1$ reaches the point $H$ in the time interval 
$\Delta t_e$. 
After $\Delta t_e$ the extreme $B_1$ has reached the position $B_2$, 
and by this time, photons emitted by the other extreme of the bar can reach
the observer simultaneously with the photons emitted by $A_1$, since
the travel paths are equal. 
The length $B_1B_2=\beta c \Delta t_e$, while $A_1H=c\Delta t_e$.
Therefore  
\begin{equation}
A_1H \, =\, A_1 B_2 \cos\theta \, \to \Delta t_e \, =\, 
{\ell^\prime \cos\theta \over \Gamma (1-\beta\cos\theta)}.
\end{equation}
Note the appearance of the term $\delta=1/[\Gamma(1-\beta\cos\theta)]$ in 
the transformation:
this accounts for both the relativistic length contraction $(1/\Gamma)$,
and the Doppler effect $[1/(1-\beta\cos\theta)]$.
The length $A_1B_2$ is then given by 
\begin{equation}
A_1B_2\, =\, {A_1H\over \cos\theta} \, =\,  
{\ell^\prime \over \Gamma(1-\beta\cos\theta)} \, =\, \delta \ell^\prime. 
\end{equation}
In a real picture, we would see the projection of $A_1B_2$, i.e.:
\begin{equation}
HB_2 \, =\,  A_1B_2\sin\theta \, =\, 
\ell^\prime\,
{\sin\theta \over \Gamma(1-\beta\cos\theta)} \, =\,  
\ell^\prime \delta \sin\theta, 
\end{equation}
The maximum observed length is $\ell^\prime$ for $\cos\theta=\beta$.

\subsection{The moving square}

Now consider a square of size $\ell^\prime$ in the comoving frame, 
moving at $90^\circ$ to the line of sight (Fig. 3).
Photons emitted in $A$, $B$, $C$ and $D$ have to arrive to the film plate 
at the same time. 
But the paths of photons from $C$ and $D$ are longer $\to$ they have to 
be emitted earlier than photons from $A$ and $B$:
when photons from $C$ and $D$ were emitted, the square was in another position.
\begin{figure*}
\vskip -1 true cm
\begin{tabular}{l l}
\hskip -1 truecm \psfig{file=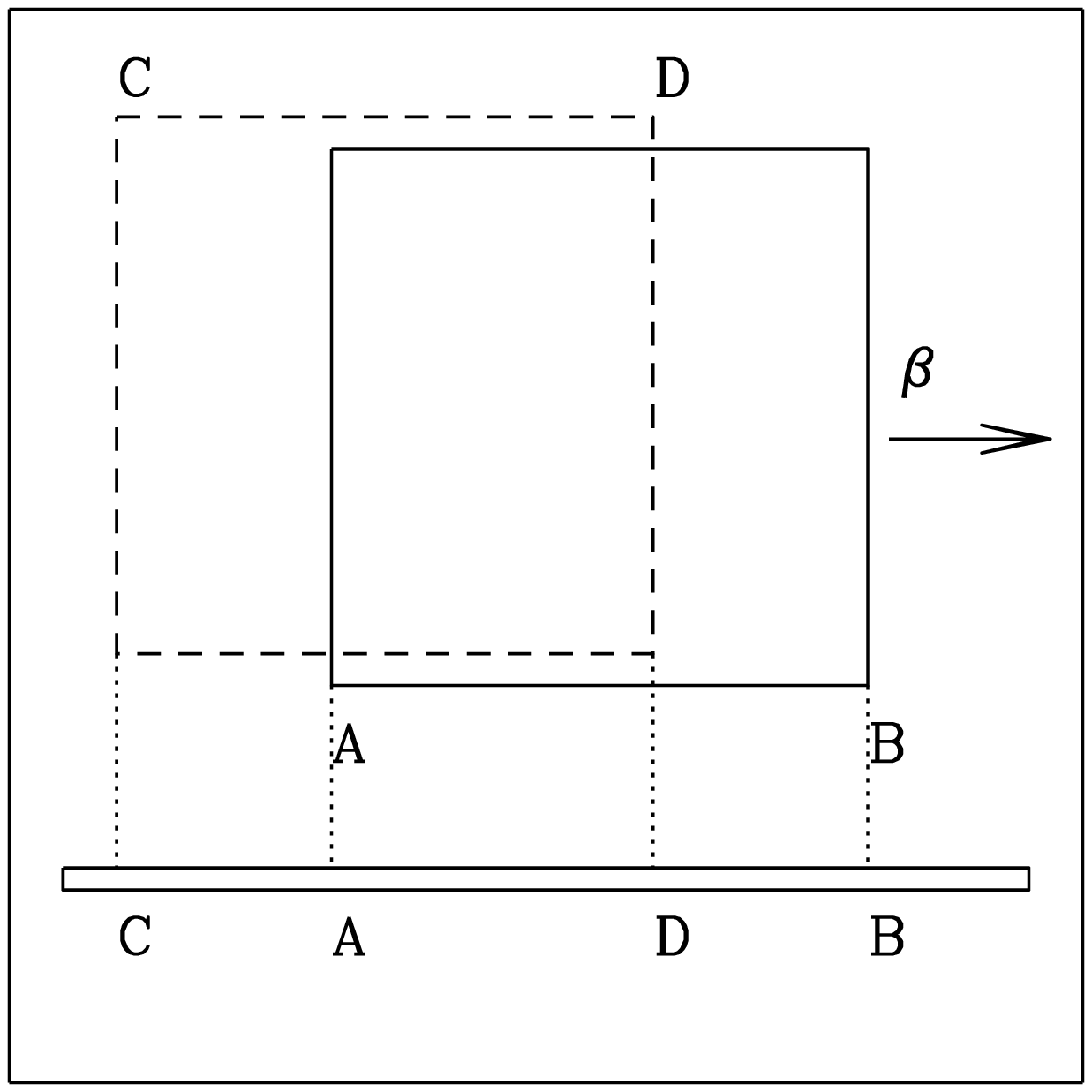,width=9truecm,height=9truecm}
&\hskip -2 truecm \psfig{file=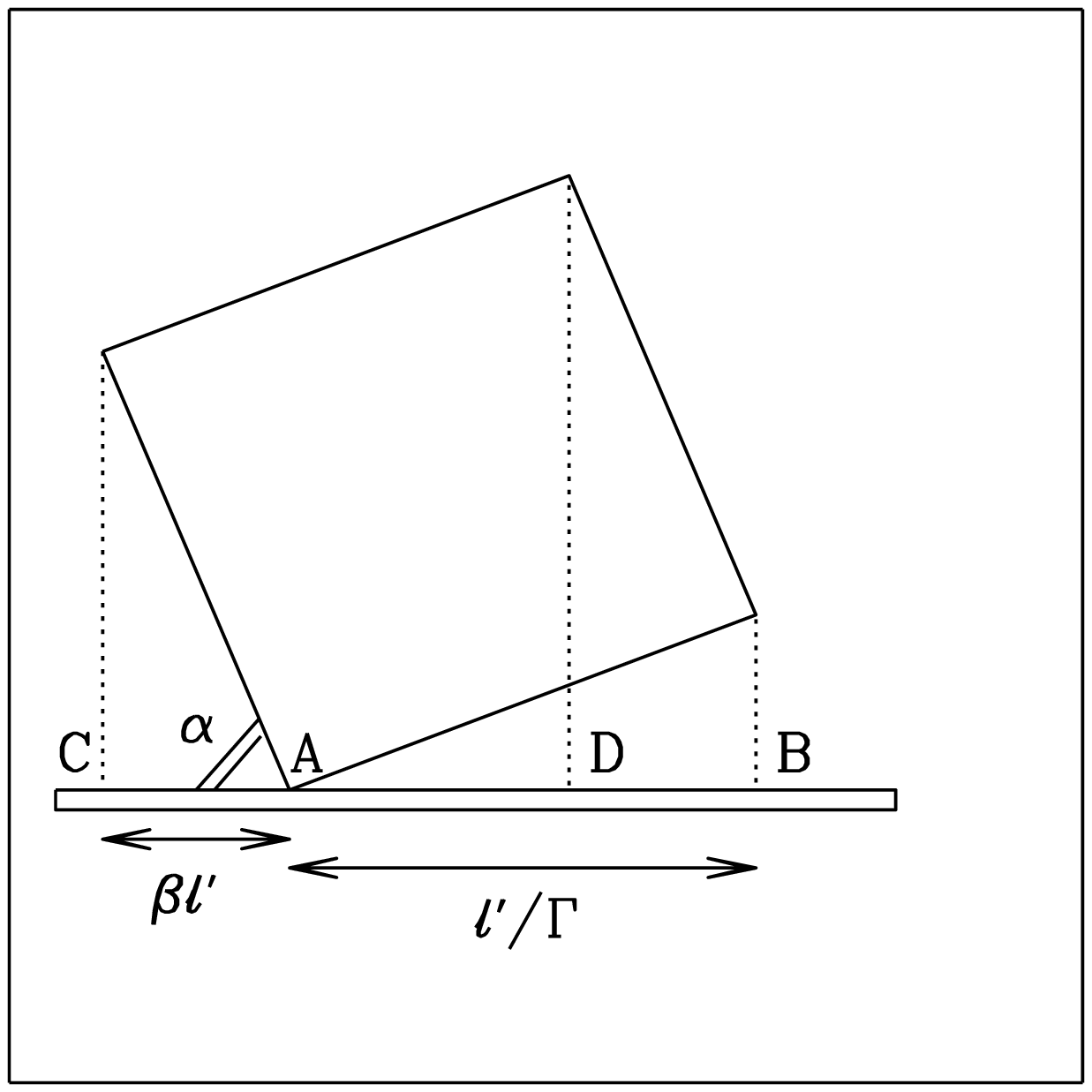,width=9truecm,height=9truecm}
\end{tabular}
\vskip -2 true cm
\caption[h]{{\it Left:}
A square moving with velocity $\beta c$ seen at 90$^\circ$.
The observer can see the left side (segment $CA$). 
Light rays are assumed to be parallel, i.e. the square is assumed to
be at large distance from the observer.
{\it Right:} The moving square is seen as {\it rotated} 
by an angle $\alpha$ given by $\cos\alpha=\beta$.}
\end{figure*}
The interval of time between emission from $C$ and from $A$ is  
$\ell^\prime/c$. 
During this time the square moves by $\beta \ell^\prime$, i.e.  
the length $CA$.
Photons from $A$ and $B$ are emitted and received at the same time
and therefore $AB=\ell^\prime/\Gamma$.
The total observed length is given by  
\begin{equation}
CB\, =\, CA+AB\, =\, {\ell^\prime\over \Gamma} \, (1+\Gamma\beta). 
\end{equation}
As $\beta$ increases, the observer sees the side $AB$ increasingly
shortened by the Lorentz contraction, but at the same time the length
of the side $CA$ increases.
The maximum total length is observed for $\beta=1/\sqrt{2}$,
corresponding to $\Gamma=\sqrt{2}$ and to $CB=\ell^\prime\sqrt{2}$,
i.e. equal to {\it the diagonal} of the square.
Note that we have considered the square (and the bar in the previous section)
to be at large distances from the observer, so that the emitted light rays
are all parallel.
If the object is near to the observer, we must take into account that different
points of one side of the square (e.g. the side $AB$ in Fig. 3) have different 
travel paths to reach the observer, producing additional distortions.
See Mook and Vargish (1987) for some interesting illustrations.

\subsection{Rotation, not contraction}

The net result (taking into account {\bf both} the length contraction 
{\bf and} the different paths) is an apparent {\bf rotation} of the square,
as shown in Fig. 3 (right panel).
The rotation angle $\alpha$ can be simply derived (even geometrically)
and is given by
\begin{equation}
\cos\alpha \, =\, \beta
\end{equation}
A few considerations follow:
\begin{itemize}
\item If you rotate a sphere you still get a sphere:  
you {\bf do not} observe a contracted sphere.
\item The total length of the projected square,
appearing on the film, is $\ell^\prime (\beta +1/\Gamma)$.
It is maximum when the ``rotation angle" $\alpha=45^\circ \to
\beta=1/\sqrt{2} \to \Gamma=\sqrt{2}$.
This corresponds to the diagonal.
\item The appearance of the square {\it is the same as what 
seen in a comoving frame for a line of sight making an angle
$\alpha^\prime$ with respect to the velocity vector,
where $\alpha^\prime$ is the aberrated angle} given by
\begin{equation}
\sin\alpha^\prime \, =\, {\sin\alpha \over \Gamma (1-\beta\cos\alpha)}\, =\,
\delta \sin\alpha
\end{equation}
See Fig. 4 for a schematic illustration.
\end{itemize}
The last point is particularly important, because it introduces
a great simplification in calculating not only the appearance of
bodies with a complex shape
but also the light curves of varying objects.

\begin{figure}
\vskip -1 true cm
\psfig{file=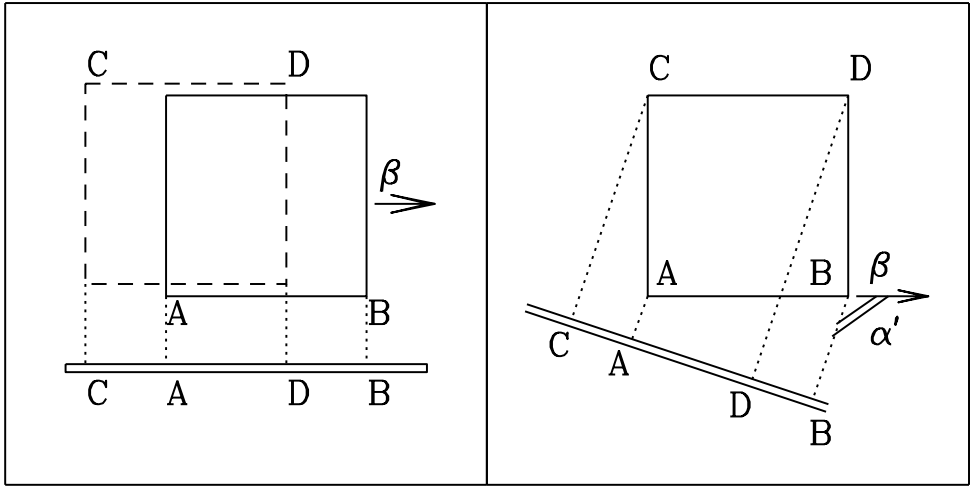,width=15truecm,height=15truecm}
\vskip -7 true cm
\caption[h]{An observer that sees the object at rest at a viewing
angle given by $\sin\alpha^\prime = \delta \sin\alpha$, will take
the same picture as the observer that sees the object moving
and making an angle $\alpha$ with his/her line of sight.
Note that $\sin\alpha^\prime=\sin(2\pi-\alpha^\prime)$.}
\end{figure}

\subsection{Light curves}
We have already seen how intrinsic time intervals $\Delta t_e^\prime$ transform
in observed $\Delta t_a$ when taking into account different photon
travel paths.
The Doppler effect can oppose to time expansion and, depending on the
viewing angle, $\Delta t_e$ can be longer or shorter than $\Delta t^\prime_e$.
There are however more complex cases, where it may be difficult
to derive a prescription as simple as Equation (2).
For instance, a relativistically moving blob which also
expands relativistically (i.e. ``a bomb" exploding in flight).
Accounting for the superposition of the two motions is complex, but
the introduction of the ``aberrated angle observer" greatly
simplifies this kind of problems:
this observer would see the blob without bulk motion and the different
travel paths are the geometric ones.
Then the observer in the lab--frame $K$ just sees the same light curve
as the ``aberrated angle observer", but with time intervals divided by $\delta$ 
and specific intensities multiplied by $\delta^3$.
In fact, in the frame $K^\prime$ the photons emitted at the
``de--aberrated angle" $\alpha^\prime$ are the very same ones
that reach the observer in $K$, at the ``aberrated angle" $\alpha$.
 
\begin{table}[h]
\hspace{1.5cm} %if you want to center your table act on this argument
\caption{Useful transformation}
\begin{tabular}{ll}
& \\
\hline
\hline
& \\
$\nu = \delta \nu^\prime$               &frequency \\
$t = t^\prime/\delta$                   &time \\
$V = \delta V^\prime$                    &volume  \\
$\sin\theta =\sin\theta^\prime/\delta$   &sine  \\
$\cos\theta =(\cos\theta^\prime+\beta)/(1+\beta\cos\theta^\prime)$  &cosine  \\
$I(\nu) = \delta^3 I^\prime(\nu^\prime)$  &specific intensity  \\
$I = \delta^4 I^\prime$                  &total intensity  \\
$j(\nu) = \delta^2 j^\prime(\nu^\prime)$  &specific emissivity  \\
$\kappa(\nu) = \kappa^\prime(\nu^\prime)/\delta$  &absorption coefficient  \\
$T_B = \delta T^\prime_B$            &brightness temperature (size directly measured)  \\
$T_B = \delta^3 T^\prime_B$          &brightness temperature (size from variability)  \\
& \\
\hline
\hline
\end{tabular}
\end{table}

\section{Conclusions}
In the last 25 years special relativity has become necessary for the 
understanding of some of the most violent phenomena in our universe,
in which large masses are accelerated to relativistic velocities.
Instead of dealing with elementary particles in accelerators or with
energetic cosmic rays, we can see fractions of a solar mass moving at 0.999$c$.
These motions give raise to several effects, the most important
being probably the collimation of the radiation along the 
direction of motion, making the source visible to distant observers.
This implies that relativistic emitting sources -- such as the plasma in 
jets of AGNs and gamma--ray bursts -- can be good probes of the far universe.

As discussed in this contribution, spatial and temporal information are 
carried by photons, and therefore the differences in their paths to
reach the observer must be taken into account.
Extended moving objects are seen rotated and therefore spheres 
remain spheres.
\vskip 0.5 true cm
\noindent
{\bf Acknowledgments}

\noindent It is a pleasure to thank Annalisa Celotti, Laura Maraschi,
Aldo Treves and Meg Urry for years of fruitful collaboration.

\end{document}